# Coherent Optical Spin Hall Transport for Spin-optronics at Room Temperature

*Ying Shi,[1,2] Yusong Gan,[1,2] Yuzhong Chen,[2] Yubin Wang,[1,2] Sanjib Ghosh,[2,*] Alexey Kavokin[3,4,*] and Qihua Xiong[1,2,5,6,*]*

[1]*State Key Laboratory of Low-Dimensional Quantum Physics and Department of Physics,*
*Tsinghua University, Beijing,100084, P.R. China.*
[2]*Beijing Academy of Quantum Information Sciences, Beijing 100193, P.R. China.*
[3]*Key Laboratory for Quantum Materials of Zhejiang Province, School of Science,*
*Westlake University, 18 Shilongshan Road, Hangzhou 310024, Zhejiang Province, P.R. China*
[4]*Institute of Natural Sciences, Westlake Institute for Advanced Study, 18 Shilongshan Road, Hangzhou 310024,*
*Zhejiang Province, P.R. China.*
[5]*Frontier Science Center for Quantum Information, Beijing, 100084, P.R. China.*
[6]*Collaborative Innovation Center of Quantum Matter, Beijing, 100084, P.R. China.*

*To whom correspondence should be addressed. Email: Q.X.: qihua_xiong@tsinghua.edu.cn; S.G.: sanjibghosh@baqis.ac.cn; A.K.: a.kavokin@westlake.edu.cn.

**Abstract:** Spin or valley degrees of freedom in condensed matter have been proposed as efficient information carriers towards next generation spintronics[1,2]. It is therefore crucial to develop effective strategies to generate and control spin or valley-locked spin currents, *e.g.*, by exploiting the spin Hall[3,4] or valley Hall effects[5,6]. However, the scattering, and rapid dephasing of electrons pose major challenges to achieve macroscopic coherent spin currents and realistic spintronic or valleytronic devices, specifically at room temperature, where strong thermal fluctuations could further obscure the spin flow. Exciton polaritons in semiconductor microcavities being the quantum superposition of excitons and photons[7], are believed to be promising platforms for spin-dependent optoelectronic or, in short, spin-optronic devices[8]. Long-range spin current flows of exciton polaritons may be controlled through the optical spin Hall effect[9]. However, this effect could neither be unequivocally observed at room temperature nor be exploited for realistic polariton spintronic devices due to the presence of strong thermal fluctuations or large linear spin splittings.[10,11] Here, we report the observation of room temperature optical spin Hall effect of exciton polaritons with the spin current flow over a distance as large as 60 μm in a hybrid organic-inorganic $FAPbBr_3$ perovskite microcavity. We show direct evidence of the long-range coherence at room temperature in the flow of exciton polaritons, and the spin current carried by them. By harnessing the long-range spin-Hall transport of exciton polaritons, we have demonstrated two novel room temperature polaritonic devices, namely the NOT gate and the spin-polarized beam splitter, advancing the frontier of room-temperature polaritonics in perovskite microcavities.

The generation and control of spin currents based on the spin Hall[12] or valley Hall effects[13], are the cornerstones of spintronics[14,15]. Despite successful generation, spin current flow over a macroscopic scale is exceedingly challenging due to rapid dephasing and spin-flips stemming from the strong electron scattering. Even more so at room temperature where thermally accelerated spin diffusion could further intensify the spin relaxation[16-18]. Thus, alternative systems or mechanism are pursued which can suppress these effects for the realization of long-range coherent spin current and realistic spintronic devices at room temperature.

Exciton polaritons, the quantum superposition of excitons and cavity photon in semiconductor microcavities[7,19,20], are believed to be promising for the room temperature spin transport over macroscopic length scales. Exciton polaritons acquire the spin degree of freedom from the photoactive coupling between excitons and photons[21], both of which have the intrinsic properties such as the spin angular momentum of an exciton and the polarization of light which manifests itself as the spin of photons in the quantum domain. However, compared to electrons, exciton polaritons as neutral particles can strongly suppress charge scattering caused by Coulomb interactions, which leads to much longer coherent lengths and lower scattering rates, so they are believed to be ideal candidates as spin current carriers[22]. Aided by the optical spin-orbit coupling[23-25], indeed the spin Hall effect of exciton polaritons was theoretically predicted[9] and soon observed at liquid helium temperatures[26,27]. However, the inability of excitons to withstand strong thermal fluctuations at room temperature rules out the possibility to create stable superfluids in traditional GaAs based semiconductor microcavities[28,29]. The microcavities based on inorganic halide perovskites do show room temperature condensates[30-32], but the presence of strong linear spin splitting originating from the intrinsic anisotropy[33-35] naturally generates a rapid spin presession and thus strongly suppresses the pure spin current flow. Owing to these outstanding challenges together with restricted short-range transports caused by imperfect growth or processing of samples, so far, no realistic spin-dependent polariton device is reported to be working at room temperature. Here, we show the unequivocal observation of the room temperature optical spin Hall effect (OSHE) of exciton polaritons using our high-quality hybrid organic-inorganic halide perovskite formamidinium lead bromide (FAPbBr$_3$) microcavity and further demonstrate the propagation of coherent spin currents over 60 μm along all directions of space. Going beyond the fundamental physics of OSHE, we realize novel room temperature polaritonic devices, such as the NOT logic gate and the spin-polarized beam splitter which nontrivially separates a linearly polarized beam into two beams of opposite spins.

Here, we consider spin-orbit coupled halide perovskite microcavities to observe coherent OSHE at room temperature and further harness it to build polariton spintronic devices. A general description of exciton polaritons in the lower polariton branch can be given by an effective Hamiltonian,

$$H = E_k + \alpha \sigma_x + \frac{\hbar}{2}\boldsymbol{\Omega}(\boldsymbol{k}) \cdot \boldsymbol{\sigma} \qquad (1),$$

where $E_k$ is the dispersion with the wavevector $k$, $\alpha$ is the linear spin-splitting due to the optical birefringence, $\boldsymbol{\Omega}(\boldsymbol{k})$ is the effective magnetic field due to the TE-TM splitting, $\sigma_{x,y}$ are the Pauli matrices and $\boldsymbol{\sigma} = (\sigma_x, \sigma_y)$ is the vector of Pauli matrices. Importantly, the Hamiltonian (1) is a characteristic of any planar microcavity, with or without excitons. Excitons are subject to an elastic scattering on a stationary disorder potential typically present in any realistic structures. The combination of elastic scattering triggered by excitons and spin-orbit interaction provided by photons results in the OSHE for exciton polaritons. Although long-range exciton polariton propagation is observed in all-inorganic perovskite microcavities, a strong linear spin splitting $\alpha \sim 8\ meV$ at $\boldsymbol{k} = 0$ hinders the realization of OSHE at room temperature[36,37]. Indeed, a large $\alpha$ manifests as a strong effective magnetic field which rotates the polariton spin between up and down states in a rapid rate and thus obscuring the flow of a pure spin current. To address this challenge, we use organic-inorganic hybrid FAPbBr$_3$ microcavity, which appears as isotropic due to the cubic crystal structure with $\alpha = 0\ meV$[38].

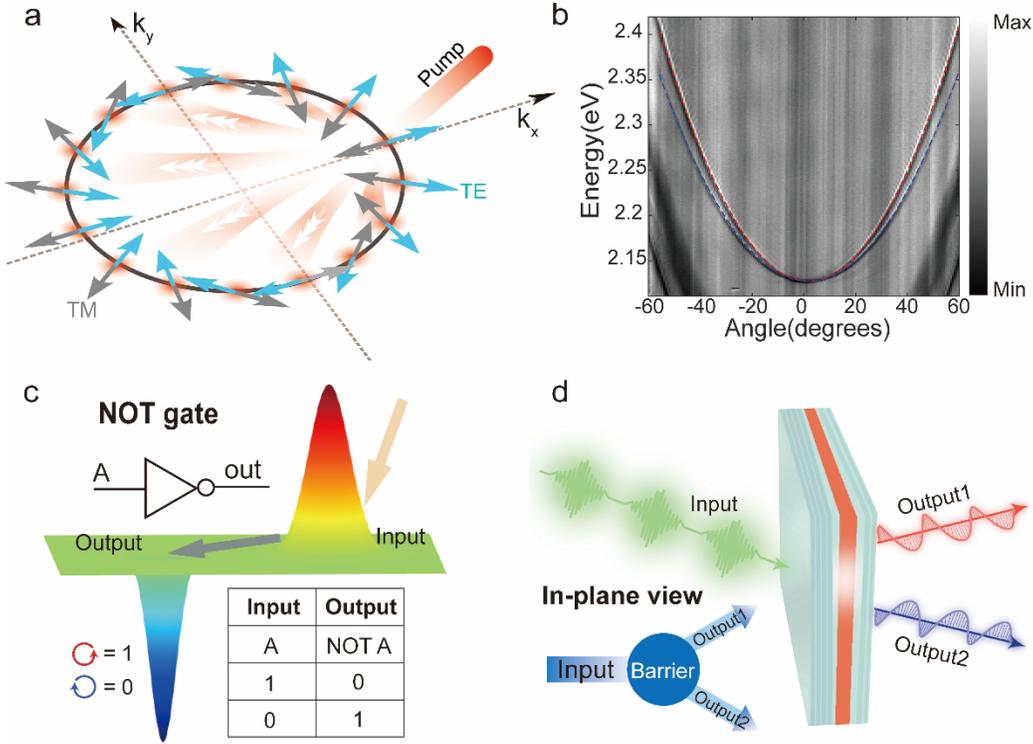

Figure 1: **The schematic diagrams and the mechanism of the optical spin Hall effect and polaritonic devices using FAPbBr$_3$ perovskite microcavity**. **a,** The illustration of the elastic scattering polariton ring generated by a linearly polarized resonant pump (the red line) and the distribution of the effective magnetic field induced by TE-TM splitting in momentum space (blue and gray arrows respectively). The effective magnetic field act on exciton polaritons to build-up a strong spin current.[55] **b**, Angle-resolved reflectivity with TE-TM splitting. Red and blue lines are the fitting results, which are obtained with the value of TE-TM splitting $\hbar\beta \sim 0.58$ meVμm$^2$. **c**, The schematic diagram of a spintronic NOT logic gate based on the propagation of polaritons in a high-quality FAPbBr$_3$ perovskite microcavity. **d**, The diagram of a spin-polarized beam splitter where the FAPbBr$_3$ perovskite microcavity device splits the input linear polarized beam into two with opposite spins.

In our system, the spin deflection is purely driven by the effective magnetic field $\mathbf{\Omega}(\boldsymbol{k})$ resulted from the TE-TM splitting of the optical microcavity, which is shown schematically in relation to the in-plane wavevector in Fig. 1a. Our sample is a planar microcavity with FAPbBr$_3$ perovskite placed at the central antinode of the confined light field (more details on sample synthesis and microcavity device fabrication can be found in Method and part I in supplementary information). A clear TE-TM splitting in an empty planar microcavity is observed by the angle-resolved reflectivity as shown in Fig. 1b, while the typical angle-resolved photoluminescence spectra of FAPbBr$_3$ perovskite microcavity are shown in supplementary information part III Fig S6-1 (along with the details in the optical setup in part II). Unlike all-inorganic perovskite microcavities, we observe no linear polarization splitting at $k = 0$ in the polariton dispersion (refer to Fig.S1a of the Supplementary Information). The observed TE-TM splitting at nonzero wavevector is well described by the effective magnetic field $\boldsymbol{\Omega}(\boldsymbol{k})$ with the components $\Omega_x(\boldsymbol{k}) = \beta(k_x^2 - k_y^2)$, $\Omega_y(\boldsymbol{k}) = 2\beta k_x k_y$. From our measured empty cavity dispersions, we find $\hbar\beta = 0.58\ meV \cdot \mu m^2$, which infuses the spin-orbit coupling in the polariton branches in our FAPbBr$_3$ microcavity.[39] We propose two functional spintronic-polaritonic devices based on the OSHE and the long-range propagation of exciton polariton. Fig. 1c shows the schematic diagram of a spintronic NOT gate based on exciton polariton spin inverting by a highly directional spin-Hall transport. Fig. 1d displays the mechanistic schematics of a spin-polarizing beam splitter, where a potential barrier breaks an incoming linearly polarized beam into two outgoing beams with opposite spins (more details can be found at the part VII of Supplementary Information).

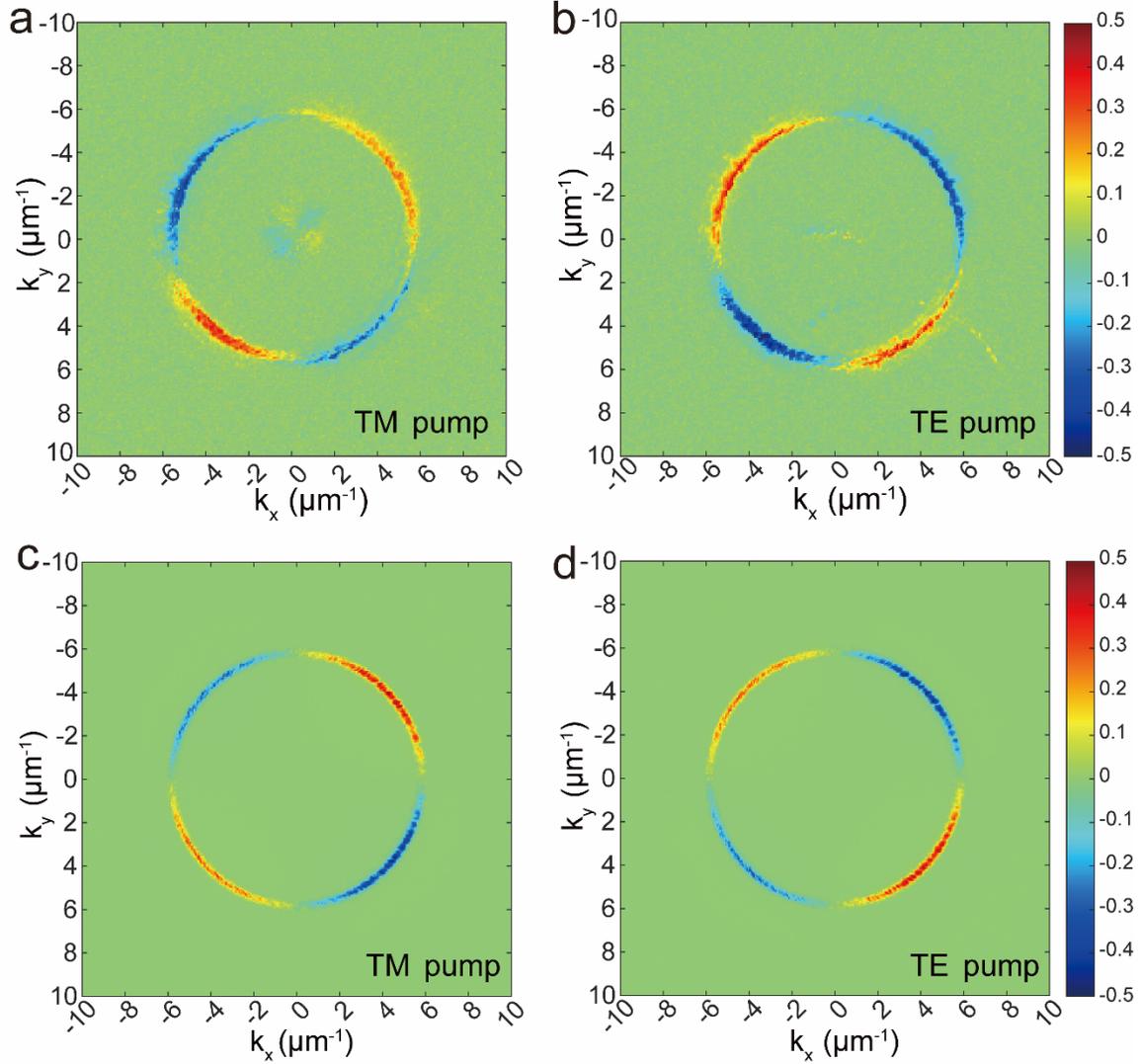

Figure 2: **Observation of the optical spin Hall effect in momentum space**. **a**, Experimental result of $s_z$ in reciprocal space under transverse magnetic (TM) polarized resonant excitation. Note that in the middle, a small spin splitting is due to the anisotropy of stress residual caused by facilitated space-confined growth. **b**, The same as a but for transverse electric polarized (TE) resonant pump. **c**, Theoretical results of a TM polarized pump. **d**, The same as **c** but for the TE polarized pump. The experimental results shown in **a** and **b** are in good agreement with theoretical results shown in **c** and **d**.

To observe OSHE at room temperature in FAPbBr$_3$ microcavities, we excited the microcavity by a 546 nm resonant linearly polarized laser with a 30° incidence angle which corresponds to a wavevector $k \sim 5.8 \, um^{-1}$ along the *x* axis. We adopted a quasi-resonant excitation in which the incident wavevector is chosen to be slightly larger compared to the wavevector given by the polariton dispersion $k = \sqrt{2mE_k/\hbar^2}$. The state corresponding to this incident angle has advantages of being slightly separated from the Rayleigh scattering ring and very far from the bare exciton resonance. Fig. 2 shows the comparative studies of the Stokes parameter $s_z$ (more details can be found in the Method section) in the momentum space between our experimental results (Fig. 2a and 2b) and theoretical simulations (Fig. 2c and 2d), which reveals the separation of spin-polarized polaritons. The observed ring-shaped distributions of polaritons in our experiments displayed by Fig. 2a and 2b represent the equal energy circle $E_k = \hbar^2 \bm{k}^2/2m$ in the momentum space. The initial direction of the wavevector $\bm{k}$ changes randomly due to the disorder (Rayleigh) scatterings from the DBR surface roughness and inhomogeneity in the low-density regime. However, the strong spin-orbit coupling in our FAPbBr$_3$ microcavity evolves the polariton spin state along the elastic scattering ring, which is primarily governed by the effective magnetic field $\bm{\Omega}(\bm{k})$[9,26,40]. Here, Fig. 2a shows the first and third quadrants of momentum space appear with positive $s_z$, while the sign reverses in the second and forth quadrants for TM excitation. This spin pattern is directly related to the periodic sign changes of $\bm{\Omega}(\bm{k})$ in the adjacent quadrants as shown in Fig. 1a. Experimentally, the maxima of the degree of circular polarization appear at angles $\pm\frac{\pi}{4}$ and $\pm\frac{3\pi}{4}$ with a magnitude as large as 0.48±0.02. Fig. 2b shows the reversal of the observed spin direction due to the rotation of the polarization plane of the incident beam (TE excitation). We find excellent agreement of our experimental results with the theoretical results shown in Fig. 2c and 2d, based on an effective two-component driven-dissipative Schrödinger's equation (refer to part VI of the Supplementary Information).

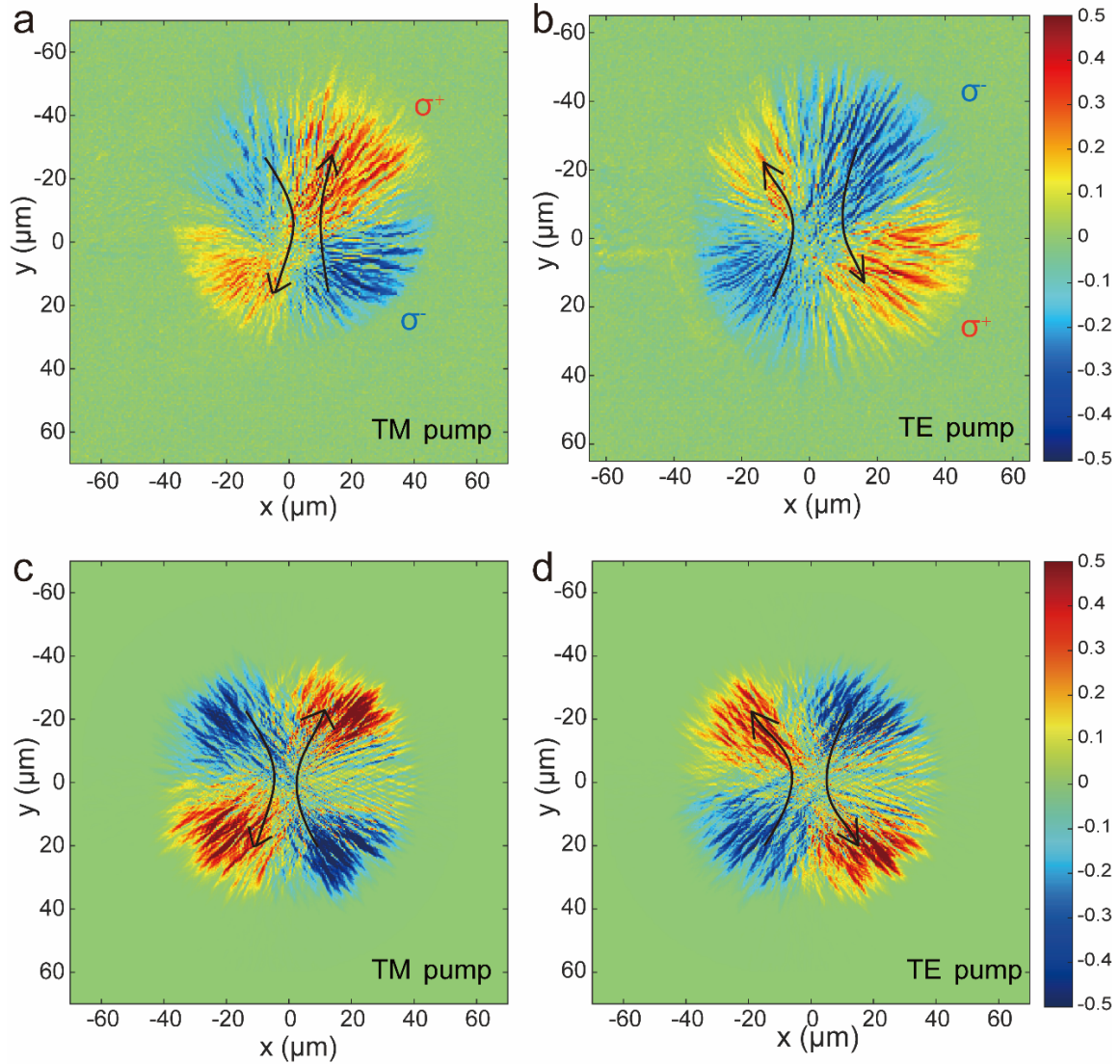

Figure 3: **Exciton polariton spin currents in real space and comparison with the theory. a, b,** Experimental Stokes parameter $s_z$ obtained from the real space circularly polarized photoluminescence under TM and TE resonant excitation. **c, d,** Theoretical results for $s_z$ under TE and TM excitations obtained by solving the driven-dissipative Schrodinger's equation. Once polaritons propagate out of the center of the pump spot, their polarizations depend on the directions of their wavevector due to OSHE. The black arrows represent directions of spin currents. Details of parameters are indicated in the Method section.

As polaritons propagate ballistically along the directions of the wavevectors on the isotropic sample plane driven by the strong nonlinear interactions, their spin deflects coherently along the effective magnetic field, resulting distinct spin patterns in real space. Fig. 3a shows clear circular polarization patterns signifying the spatial separation of polaritons with spin-up and spin-down. Fig. 3b shows the change in the spin direction when the pump laser is changed from TM to TE polarizations. We observe polariton propagation nearly 60 $\mu m$ carrying the spin information, which indicates the generation of a strong spin current. The images presented in Figures 2 and 3 are very similar to the experimental signatures of the optical spin Hall effect obtained in a GaAs based microcavity at the cryogenic temperature[26]. Our observation is fully supported by the theoretical results shown in Fig. 3c and 3d (more information can be found in part VI of the Supplementary Information).

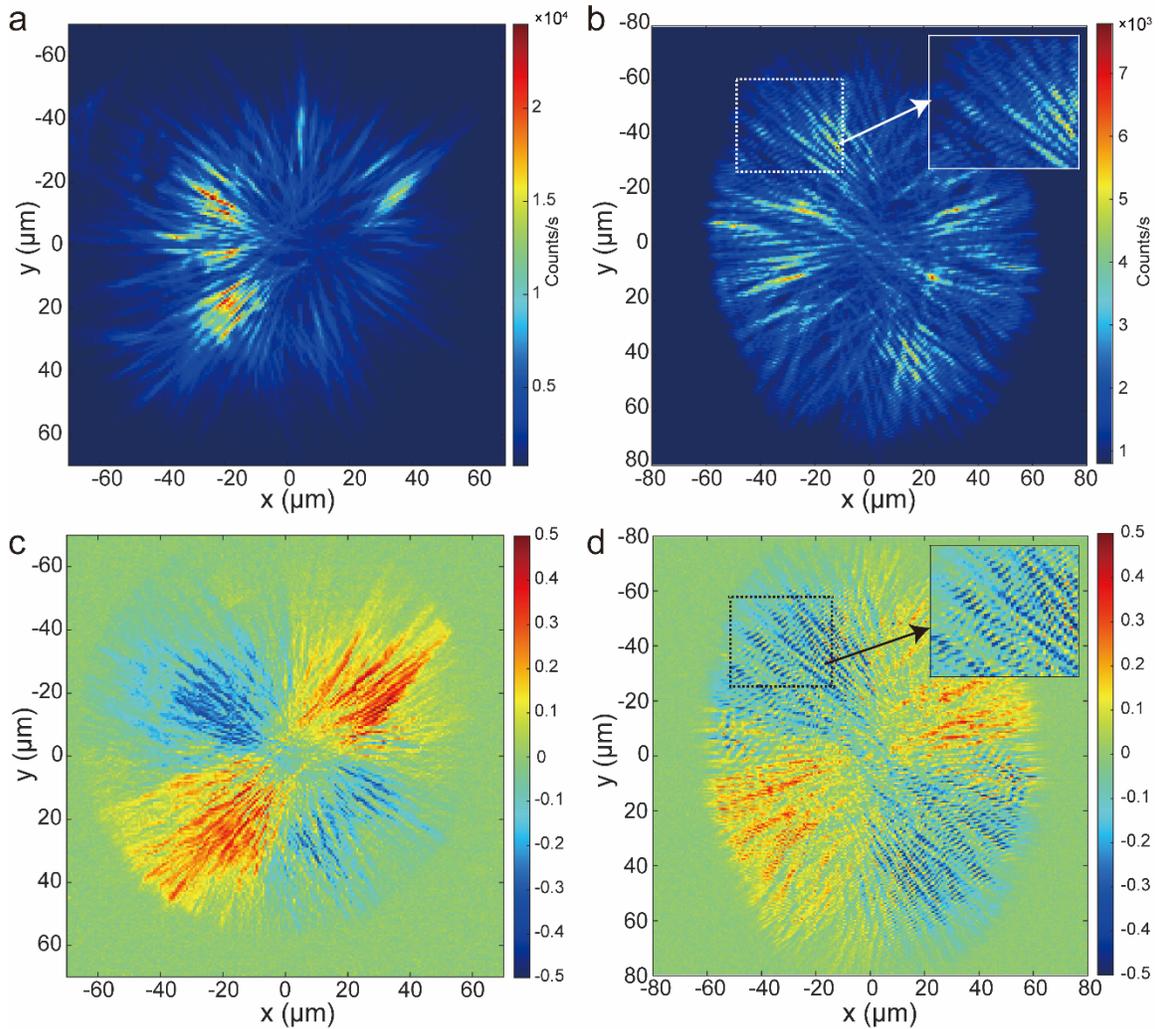

Figure 4: **Characterization of the exciton polariton long-range coherence**. **a**, The real space total intensity under a resonant TM pump. **b**, The superposition of the real space image and its inverted image from two arms of Michelson interferometer. The upper right inset is a zoom-in view of the dashed white

box. Clear interference fringes can be observed in specific directions up to a distance as large as 60μm, which demonstrates the spatial coherence during exciton polariton propagation. **c**, Measured real space circular polarization degree $s_z$ under a resonant TM polarized excitation. **d**, The superimposed interference fringes obtained separately for the right and left circularly polarized photoluminescence (PL). The inset is a zoom-in view of the dashed black box. The coherent polariton spins during propagation are measured by selectivity detecting the emission intensity of right and left circularly polarized PL.

Ballistic propagation of exciton polaritons does not automatically guarantee the coherence, which could be lost due to noise. To prove the coherence, we directly performed interferometric measurements based on the Michelson interferometry for the first-order spatial coherence $g^{(1)}(\mathbf{r}, -\mathbf{r})$, which signify the phase coherence between the points **r** and **-r** in real space.[41] The real-space image shown in Fig. 4a was sent to the Michelson interferometer with one arm replaced by a retroreflector, which allows centrosymmetric image inversion. Fig. 4b presents the interference fringes throughout the whole region of polariton propagation which spans over a distance as large as 60 μm. The clear interference fringes are observed because the inherited coherence from the excitation laser wasn't lost due to relaxation during the ballistic propagation within their radiation lifetime.[21] Our results unambiguously demonstrate the long-range spatial coherence of exciton polariton propagations in our FAPbBr$_3$ perovskite microcavity system at room temperature. Moreover, to establish the coherent spin transport along with the polariton mass propagation, we performed the coherence measurement with the photoluminescence (PL) obtained in two circular polarization bases. Fig. 4c shows the two circular polarization signals superimposed in the same figure with opposite sign and normalized by the total intensity. Fig. 4d shows the corresponding interference fringes for the spin-polarized PL which clearly shows the coherence in the spin transport throughout the entire region where polaritons have propagated along all directions.

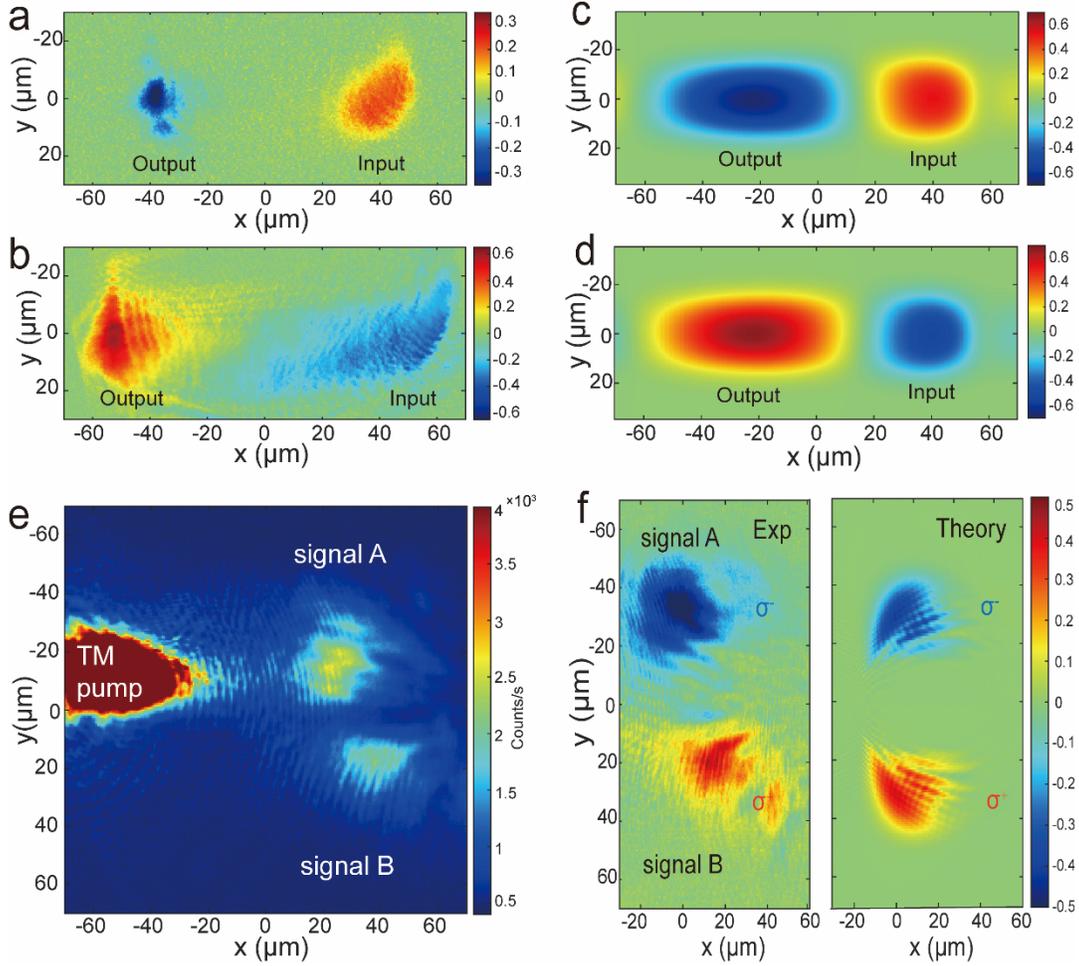

Figure 5: **Realizing polariton spintronic devices by harnessing spin Hall transport**. **a**, Experimental results of a spintronic NOT gate for right circular polarized input. **b**, Operation of the NOT gate for left circular polarized input. Clearly the outputs in **a** and **b** are both inverted to the opposite spin polarization with respect to the input. **c** and **d** show theoretical simulations corresponding to the experimental results **a** and **b** respectively. **e**, The real space PL from the spin polarizing beam splitter where the incoming linearly polarized beam is separated into two outgoing beams. **f**, Experimental and theoretical degrees of circular polarization of the outgoing beams of the polarizing beam splitter device.

In our system, we have established OSHE and demonstrate long distance coherent propagation. Using the principle of OSHE and properties of long-range propagation properties, we design two room temperature spintronic-polaritonic devices. The TE-TM splitting can make a full spin rotation of polariton from right circular to left circular polarization in microcavity with long propagation distance.[42,43] Fig. 5a and 5b show the operation of a spintronic NOT gate, where the incoming spin is inverted to the opposite spin. Our mechanism for the spin inversion is based on the highly directional spin Hall transport (The schematic diagram is shown in Fig. 1c), which was possible due to scattering free propagation of exciton polaritons in our high-quality FAPbBr3 microcavity at room temperature. The unidirectional propagation is described by the initial wavevector $k_x = 5.8\ um^{-1}$ and $k_y = 0$ which corresponds to $\Omega_x = \hbar\beta k_x^2$ and $\Omega_y = 0$. Under such an effective magnetic field, the polariton spin state periodically oscillates with the propagation distance (see the part VII of Supplementary Information S12). Thus, an appropriate choice of the propagation distance can invert the initial spin state as can be seen in Fig. 5c and 5d, which agrees well with the experimental results. Furthermore, we have devised a spin-polarized beam splitter, which separates an incoming linear polarized beam into two oppositely circular-polarized beams. Here, we utilize a naturally occurred barrier to separate the incoming beam into two as shown in Fig. 5e (refer to Fig.S12b and Fig.S13a in the Supplementary Information). In a trivial system, the spin polarization of outgoing beams would be the same as the incoming one. However, in our system the strong spin-orbit coupling imposes a spin rotation. Moreover, the spin rotation is determined by the effective magnetic field configuration shown in Fig. 1a. By meticulously choosing the splitting angles, we achieved opposite spin-polarizations for the outgoing beams as can be seen in Fig. 5f. Our simulated results, based on a barrier potential and a spin-orbit coupled Hamiltonian, fully agrees with the experimental demonstrations at room temperature (more details can be referred to part VII section 13). In the future, one can use electron beam lithography technique and a controllable magnetic field to design the controlled-NOT gate.[44] Moreover, by introducing two control light beams, one can realize other logic gate operations[45,46] for achieving all-optical logic operable on-chip circuitry[42,47].

In conclusion, we have unambiguously observed coherent optical spin Hall effect at room temperature in an organic-inorganic FAPbBr3 perovskite planar microcavity, with a hallmark feature of the spatial separation of polaritons with different spins. Our work demonstrates the capability of generating and controlling the exciton polariton spin current, *e.g.*, the polariton spin propagation direction can be changed by changing the polarization of the incident light. We have shown that the exciton polariton spin preserves the long-range spatial coherence during the propagation, based on which we have demonstrated polariton spin-optronic NOT gate and a nontrivial spin-polarizing beam splitter. Our observation of the room temperature coherent spin Hall effect and harnessing it for the spin-optronic devices advocate the exceptional promise towards room temperature spin-based computing and information processing.

**Materials and Methods**

**Microcavity Fabrication**

Hybrid organic-inorganic halide perovskite FAPbBr$_3$ single crystal with sub-cm size and controllable thickness was directly grown on transparent sapphire substrate with the bottom distributed Bragg reflector (DBR), which is consisted of 15.5 pairs of silicon dioxide and titanium dioxide deposited by electron-beam evaporation, by diffusion facilitated spatial restricted growth method[48]. A thin layer of PMMA was then spin-coated to protect the perovskite sample. Then the substrate was put into the electron-beam evaporator chamber to fabricate the top DBR (8.5 pairs of silicon dioxide and tantalum pentoxide). More details can be found in part I in supplementary information.

**Optical spectroscopy characterization**

The momentum-space and real-space spectroscopy measurements are collected by a home-built photoluminescence setup in the Fourier imaging configuration. To observe OSHE, the perovskite microcavity was resonantly pumped by a linearly polarized laser to the lower polariton branch with a picosecond pulse width, a repetition rate of 0.5 MHz, and an angle of incidence ~30° (a supercontinuum laser SC-Pro and optoacoustic tunable filter AOTF-Pro by YSL photonics was used). The pump laser, which is slightly elliptic, was focused down to a ~30 μm diameter spot and collected by 50× microscope objective (NA=0.8) covering an angular range of $\pm 53.1°$. The excited light can be filtered out by a customized baffle in the collection path due to a large angle of incidence. The linear-polarized PL can be measured with a quarter waveplate, a half waveplate and a linear polarizer. To observe exciton polariton condensation, the FAPbBr$_3$ perovskite microcavity was pumped by non-resonant excitation centered at 430 nm with a pulse duration of 100 fs and a repetition rate of 5 kHz (light conversion optical parametric amplifier Topas Prime pumped by a Coherent Ti-Sapphire regenerative amplifier Astrella USP 5kHz). The excitation laser beam was focused down to a ~24 μm diameter spot on the sample. For a continuous wave laser of 405 nm pumping and white light illumination, the angle-resolved PL and reflectivity were measured through a high numerical aperture 100× microscope objective (NA=0.9), covering an angular range of $\pm 64.1°$. The signal was sent to a 550 mm focal length spectrometer (Horiba iHR550) with a 600 gr/mm grating and a 256 × 1024 pixel liquid nitrogen cooled charge-coupled device (CCD). The complete description of the set-up is shown in Fig. S9 in supplementary information.

**Stokes parameters**

The polarization of the resonant excitation laser (which can be either TE or TM polarization) can be injected directly into the initial spin state of polaritons. In photoluminescence measurement, the emitted light carries the spin state information of polaritons through its polarization, which is characterized by the Stokes vector $\mathbf{S} = (s_x, s_y, s_z)$ with the components given by,

$$s_{x,y,z} = \frac{I_{H,D,\circlearrowright} - I_{V,A,\circlearrowleft}}{I_{H,D,\circlearrowright} + I_{V,A,\circlearrowleft}},$$

$I_{H/V,D/A,\circlearrowright/\circlearrowleft}$ are the intensities collected in the horizontal or vertical, diagonal or antidiagonal, and right or left circular polarization measurement basis respectively.

**Theoretical calculation**

We consider the circular polarization components of wave function as the calculation basis, which is given by $\Psi_\sigma(x,y,t)$, where σ could be + or −. Our simulation is based on a two components driven-dissipative Schrödinger's equation:

$$i\hbar \frac{d}{dt}\Psi_\sigma(x,y,t)$$
$$= \left(-\frac{\hbar^2 \nabla^2}{2m} - i\Gamma(x,y) + V(x,y)\right)\Psi_\sigma(x,y,t)$$
$$+ \frac{\hbar}{2}\beta\left(i\frac{\partial}{\partial x} + \sigma\frac{\partial}{\partial y}\right)^2 \Psi_{-\sigma}(x,y,t) + iF_\sigma(x,y,t),$$

$V(x,y)$ is a random disorder potential with certain amplitude and correlation length.[49] The term $\frac{\hbar\beta}{2}\left(i\frac{\partial}{\partial x} + \sigma\frac{\partial}{\partial y}\right)^2$ describes the effect of TE-TM splitting which cause the optical spin Hall effect, and $F_\sigma(x,y,t)$ is the resonant optical pump with the following form:

$$F_+(x, y, t) = F_0 \cdot e^{-ik_{in}x-((x-x_c)^2+(y-y_c)^2)/L^2-t/\tau},$$

$$F_-(x, y, t) = F_0 \cdot e^{-ik_{in}x-((x-x_c)^2+(y-y_c)^2)/L^2-t/\tau}.$$

We numerically solve the equation and calculate $\Psi_\sigma(x, y, t)$. The total light intensity can be expressed by:

$$s_0(x, y, t) = |\Psi_+(x, y, t)|^2 + |\Psi_-(x, y, t)|^2.$$

These two terms also represent the intensity of left and right circularly polarized light separately. Then we calculate the degree of circular polarization (or the z component of the Stokes vector) with:

$$s_z(x, y, t) = \frac{|\Psi_+(x,y,t)|^2 - |\Psi_-(x,y,t)|^2}{s_0(x,y,t)}.$$

We can also get the *k*-space information by a two-dementional Fourier transform from $\Psi_\sigma(x, y, t)$ to $\Psi_\sigma(k_x, k_y, t)$ and similarly calculate $s_0$ and $s_z$ in momentum space. More details of numerical simulations are shown in the Supplementary Information. The parameters we used in our calculation are given by, $k_{in} = 5.8\ \mu m^{-1}$, $\hbar\beta = 0.013\ meV \cdot \mu m^2$, polariton effective mass at $k_{in}$ is $m = 5.6 \times 10^{-5} m_e$ ($m_e$ is the free electron mass), the pump spot diameter is 24 µm. The disorder potential has an r.m.s. amplitude of 1.5 meV and a correlation length of 0. 5 µm.

**Acknowledgement**

We thank Dr. Quanbing Guo for his helps during our optical experiments. Q.X. gratefully acknowledges funding support from the National Natural Science Foundation of China (Grant no. 12020101003, and 12250710126) and strong support from the State Key Laboratory of Low-Dimensional Quantum Physics at Tsinghua University. S.G. gratefully acknowledges funding support from the Excellent Young Scientists Fund Program (Overseas) of China, the National Natural Science Foundation of China (Grant No. 12274034) and the start-up grant from Beijing Academy of Quantum Information Sciences.

**Author contribution**

Q.X. and S.G. conceived the idea. Y.S. performed prepared the samples and conducted all the optical spectroscopy measurements. Y.G. performed the theoretical calculations. Y.C and Y.W. provided help on the measurements. Y.S. wrote the manuscript with the inputs from Y.G., S.G. and Q.X. A.K. provided input in theoretical understanding. All authors participated in analyzing the results and preparing the manuscript and agreed with the conclusion. Q.X. and S.G. supervised the whole project.

**Competing interests**

The authors declare no competing interests.